\documentclass[aps,prb,preprint,superscriptaddress]{revtex4}
\usepackage[dvips]{epsfig}

\begin{document}


\title{Quenching of phase coherence in quasi-one dimensional ring crystals}



\author{K. Shimatake}
\affiliation{Department of Applied Physics, Hokkaido University, Kita 13 Nishi 8 Kita-ku, Sapporo 060-8628, Japan.}
\author{Y. Toda}
\affiliation{Department of Applied Physics, Hokkaido University, Kita 13 Nishi 8 Kita-ku, Sapporo 060-8628, Japan.}
\affiliation{PRESTO Japan Science and Technology Agency, 4-1-8 Honcho, Kawaguchi, Saitama, Japan.}
\author{S. Tanda}
\affiliation{Department of Applied Physics, Hokkaido University, Kita 13 Nishi 8 Kita-ku, Sapporo 060-8628, Japan.}
\date{\today}

\begin{abstract}
The comparison of the single-particle (SP) dynamics between the whisker and ring NbSe$_3$ crystals provides new insight into the phase transition properties in quasi-one-dimensional charge density wave (CDW) systems. In the incommensurate CDW phase, the SP relaxation triggered by an ultrafast laser pulse reflects a formation of collective states, and shows a divergence of relaxation time when approaching a transition temperature. The degree of divergence is less pronounced in the ring than that in the case of the whisker, suggesting a loss of phase coherence in the ring crystal characterized by a closed-loop topology.
\end{abstract}


\maketitle
During the past decades, there has been intense interest in understanding the properties of the charge density wave (CDW) in quasi-one-dimensional (1D) metals, one of the most widely discussed issues of which is the influence of phase fluctuations on the CDW transitions \cite{Gruner}. In an uncoupled-1D system, a large fluctuation strongly decreases the CDW transition temperature ({\it T$_{c}$}) to below the mean field transition temperature since CDWs undergo a phase transition only when the 3D long-range order develops \cite{Gruner}. Indeed, a pronounced finite size effect has been observed in NbSe$_3$, in which the reduction of the number of parallel chains makes the transitions less pronounced and decreases {\it T$_{c}$} \cite{size}. In this sense, it is also expected that the crystal topology makes a substantial contribution to the phase transition since topology imposes additional constraints on inter-chain correlations. However, the effects of crystal topology on CDW properties are still poorly understood.

Recently, several 1D transition metal chalcogenides of the type MX$_3$ were found to have various types of topological crystals, where the whisker crystals naturally form the ring, M\"{o}bius, and figure-of-eight geometries \ \cite{Tanda, Tsuneta}. Because of both the small sizes and damage-free formations of these structures, it is possible for CDWs to maintain their coherence within individual chains with a closed loop, thus making it a good candidate for studying the topological effects on the long-range ordering. In this work, we have investigated the topological crystals of the NbSe$_3$ compound by measuring the time-resolved optical reflectivity changes. The comparison of single-particle (SP) decays between the whisker and ring structures reveals a significant difference, suggesting an influence of the crystal topology on their phase transitions.

Figures 1(a) and (b) show the scanning electron micrographs of the NbSe$_3$ whisker and ring, respectively. Both crystals were prepared by the chemical vapor transport method under virtually the same conditions. The whisker is a standard bulk with a length of a few mm (along the conducting axis {\it b}) and a width of 50 $\mu$m. The ring has a somewhat disklike structure, the dimensions of which are 50 $\mu$m in outer diameter and several $\mu$m in internal diameter (conducting axis {\it b} in the radial direction). The circumferential length of a center hole is around 10 $\mu$m and is comparable to the correlation length of CDWs ($\xi_{\parallel b}>2.5 \mu m$ \cite{Sweetland}) in this compound. The details of the growth mechanism for the ring and its structural analysis based on X-ray diffraction have been described in detail elsewhere \cite{Tsuneta}.

To evaluate the CDW properties, we employed an optical measurement that enables an ideal observation for topological structures since the photon can act as a noncontact probe and therefore preserve the crystal topology. The time-resolved measurement was achieved by a conventional pump-probe technique with a micro optical setup. For the excitation source, we used a mode-locked Ti:sapphire laser with a pulse width of $\sim$130 fs centered at an energy of 1.56 eV with a repetition rate of 76 MHz. We detected the reflectivity change ({\it $\Delta$R}) of the probe pulse with a certain delay from the pump pulse. The pump and probe pulses were orthogonally polarized and focused through an objective lens onto a single-crystal region. The pulses overlap with a diameter of 10 $\mu$m on the sample surface was monitored using a charge coupled device camera, and was kept at a fixed position during the measurements in each sample. For the ring sample, the position was fixed to be near the inner hole in order to emphasize its topological character. The pump and probe fluencies were $\sim$40 $\mu$J/cm$^2$ and $\sim$10 $\mu$J/cm$^2$, respectively. The steady state heating caused by the laser was accounted for by measuring the excitation-power dependence of {\it $\Delta$R}.

\begin{figure}[t]
\begin{center}
\includegraphics[width=87mm]{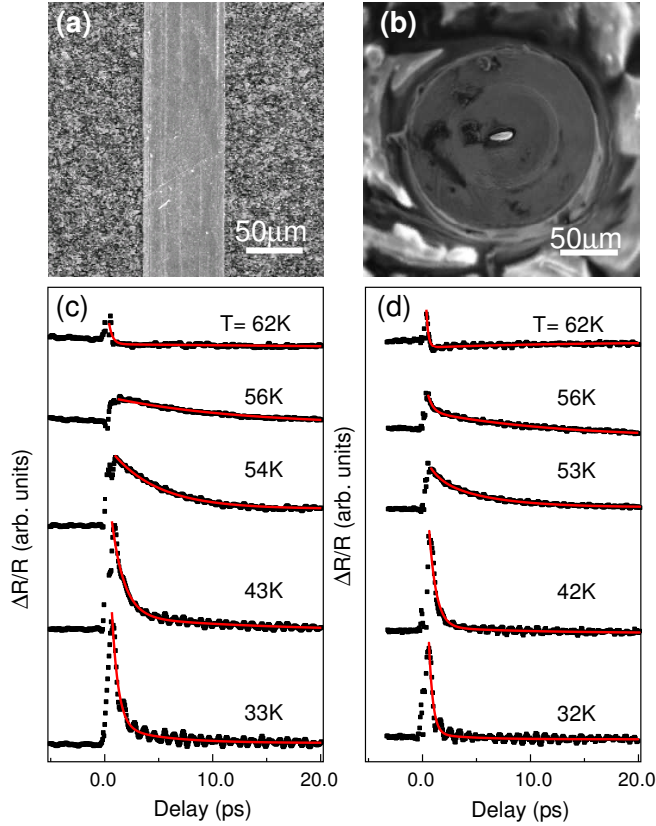}
\caption{Electron micrographs of the topological samples: (a) whisker and (b) ring. The corresponding transient reflectivity changes, {\it $\Delta$R/R}, at various temperatures below and above T$_{c2}$ in (c) and (d). The solid lines indicate the results of least-squares fitting.}
\end{center}
\end{figure}

NbSe$_3$ consists of three pairs of metallic chains parallel to the conducting axis, and exhibits two incommensurate CDWs with transition temperatures at {\it T$_{c1}$}=145 K and {\it T$_{c2}$}= 59 K. Note that similar {\it T$_c$} ({\it T$_{c1}$}=140.8 K and {\it T$_{c2}$}= 57.4 K) have been found in the ring samples by resistivity measurement \cite{Tsuneta}.
Figures 1(c) and (d) show several transient {\it $\Delta$R} of a whisker and a ring, respectively, for various temperatures below and above {\it T$_{c2}$}. As has been observed in K$_{0.3}$MoO$_3$ and several other CDW compounds \cite{Demsar, AM}, the signal at the lowest temperature is dominated by two features: a combination of exponential responses and damped sinusoidal oscillations. We assign the exponential part to the transient responses of the SPs, while the oscillation part reflects coherent motions including phonon and collective CDW modes induced by instantaneously photoexcited SPs.

We will now focus our attention on the SP dynamics. The interpretation of the transient {\it $\Delta$R} in CDW systems has been established by analogy to that in general materials such as metals and semiconductors \cite{Demsar, Kabanov2, Kabanov1}. Since the pump pulse with near-infrared energy can excite the SPs into continuum states far above the CDW gap, the relaxation down to states near the band edge results in an abrupt increase of {\it $\Delta$R}. The CDW gap then causes carriers to accumulate in its upper edge. A subsequent decay thus reflects the transient density change of these accumulated carriers. Since the CDW gap depends on the sample temperature, this fast decay exhibits temperature dependence associated with the formation of the gap \cite{Demsar, Kabanov1}. A long-lived decay can be attributed to relaxation from phason states pinned just above the ground state and/or impurity-related trapped states within the gap \cite{Kabanov2}. 

When increasing the temperature to {\it T$_{c2}$}, the decay time in both types of crystals shows an increase followed by an abrupt drop above {\it T$_{c2}$}. The relaxation process across the gap is manifested as phonon emissions and absorptions. Since the gap energy of CDWs is strongly temperature dependent and decreases on approaching {\it T$_{c2}$} from below, the increase of phonon density, which can contribute to the re-absorption process, results in an increase of the relaxation time. The divergence of decay time just below {\it T$_{c2}$} thus reflects the formation of a collective gap.

It is important to note that no significant changes of {\it $\Delta$R} occur at around {\it T$_{c1}$} under the present experimental conditions with a probe energy of 1.56 eV. On the other hand, {\it $\Delta$R} with the probe energy of 1.2 eV shows characteristic diverse behaviors at each transition temperature. In the latter case, however, the SP response from another chain can contribute more to the signal below {\it T$_{c2}$} than in the former case. In addition, {\it T$_{c1}$} is suitable for analyzing characteristic SP dynamics because the transition at lower temperature reduces the thermal fluctuation and makes the collective gap formation clearer than that at higher temperature.
\begin{figure}[t]
\begin{center}
\includegraphics[width=87mm]{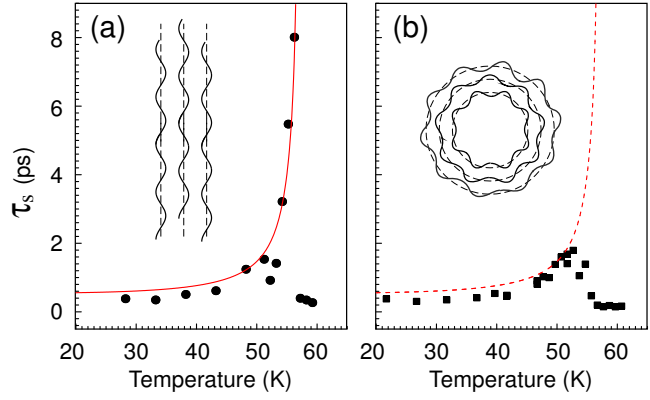}
\caption{Temperature dependences of $\tau_s$ in the ring (a) and whisker (b) around {\it T$_{c2}$}. The solid line in (a) shows a theoretical fit \cite{Kabanov1} to the data, and is also shown in (b) for comparison, where a large reduction of $\tau_s$ and disagreement with the theoretical curve are clearly seen. Insets: sketches of CDWs modulated on the neighboring chains. Below mean-field temperature, Coulomb repulsion tends to align the CDWs out of phase with each other. }
\end{center}
\end{figure}

Figures 2 (a) and (b) show plots of the fast decay time ($\tau_s$) as a function of temperature in the ring and whisker, respectively. Following the SP relaxation processes described above, we evaluated $\tau_s$ by a least-squares fitting procedure with a sum of two exponential functions. Below $\sim$40 K, $\tau_s$ in both samples are nearly constant and are identical ($<$ 1 ps). 
On the contrary, a remarkable difference is seen above T=50 K, where the $\tau_s$ diverges due to the reduction of the gap energy. However, the degrees of divergence of $\tau_s$ are very different in these two samples. For better comparison, simultaneous plots of transient signals associated with the longest decay time are represented in Fig. 3. The signal in the whisker exhibits an extremely long exponential decay, the evaluated $\tau_s$ of which reaches as long as almost 10 ps. In contrast, $\tau_s$ in the ring is only 2 ps.
\begin{figure}[h]
\begin{center}
\includegraphics[width=63mm]{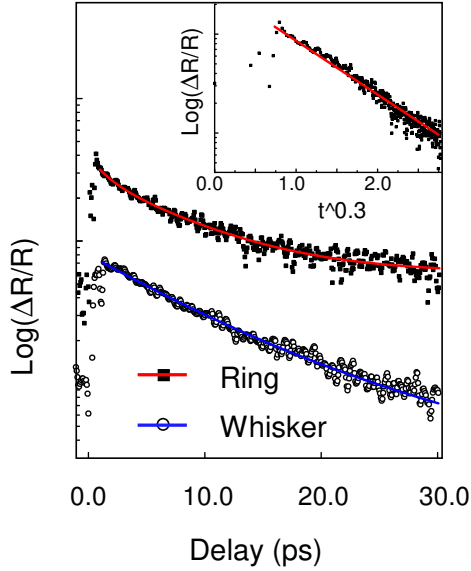}
\caption{Semi-logarithmic plots of the reflectivity changes for the whisker (T=56 K) and the ring (T=52 K), where the longest $\tau_s$ was obtained in each crystal. The plots are shifted vertically for clarify. Inset shows another plot for the ring. The solid line indicates the slope of $\exp{(-t/\tau)^{0.3}}$, where $\tau$ is 1.61.}
\end{center}
\end{figure}

For another qualitative analysis of $\tau_s$, we fit the data by a BCS-like temperature dependence of the gap (2$\Delta_g$(T)). The solid curve in Fig. 2 (a) represents the theoretical fit given by Kabanov {\it et al.}\cite{Kabanov1} to the data with 2$\Delta_g$(0) $\approx$ 50 meV. Although 2$\Delta_g$(0) in NbSe$_3$ obtained in previous experiments varies significantly \cite{STM}, 2$\Delta_g$(0) $\approx$ 50 meV obtained in our optical measurement is in good agreement with the maximum gap value of 45$\pm$10 meV in angle-resolved photoemission spectroscopy \cite{ARPES}. On the other hand, we cannot accurately fix the parameter using the data of the ring. This is due to the quenching behavior of $\tau_s$ at around {\it T$_{c2}$}. Instead, the curve optimized for the whisker is represented as the dashed line in Fig. 2 (b). In the vicinity of {\it T$_{c2}$}, a marked deviation from the theoretical fit is observable. This discrepancy suggests that the transient {\it $\Delta$R} reflects the critical Ginzuburg-Landau fluctuations, where $\Delta_g$(T) cannot be reproduced by the conventionally used theoretical analysis based on the mean-field approximation.

On the basis of the different temperature dependences of $\tau_s$ between the crystals, we now consider the topological effect on the phase transition properties.
The difference in the longest $\tau_s$ at {\it T$_{c2}$} is attributed to the difference in the coherence of the collective states between the samples. In this case, the suppressed divergence of the decay time observed in the ring indicates the quenching of the phase coherence for a long-range ordering. The insets of Figs. 2 (a) and (b) are schematic illustrations of the candidate spatial distributions of charge density along the individual chains for the whisker and the ring, respectively. Let us recall the process of phase transition in CDW materials. In the quasi-1D system, the Coulomb correlation between adjacent chains should be taken into account in the phase transition \cite{Gruner}. In the whisker, which has no boundary, 3D ordering can be easily realized by adjusting the neighboring chains with a phase difference of $\pi$. In contrast, for the ring crystal, CDWs with the same oscillation period in the closed loop reduce the Coulomb repulsion only by undergoing a lattice expansion, resulting in a loss of phase coherence between adjacent chains. This topologically induced phase fluctuation is also consistent with deviations from the BCS-like temperature dependence in the vicinity of {\it T$_{c2}$}. The enhanced phase fluctuation inhibits the divergence of the relaxation time represented by the mean-field approximation. As a result, the transient {\it $\Delta$R} in the ring renders the critical phase fluctuations clear at around {\it T$_{c2}$}. So far, only the size effect has been focused on in discussions of the inter-chain interaction in quasi-1D systems \cite{size}. However, based on the mechanism discussed above, crystal topology can also enhance the phase fluctuations between the neighboring chains. 

We now briefly comment on the relaxation function of the transient $\Delta R$ in the vicinity of {\it T$_{c2}$}. Recently, Nogawa {\it et al.} proposed a phase field model for CDWs in ring crystals remarking on the frustration between intra- and inter-chain couplings \cite{Nogawa}. By using Monte Carlo simulations, they found that the relaxation function in the low temperature ordered phase obeys a power law decay instead of the usual exponential decay. On the other hand, we should note that the transient $\Delta R$ for the ring is well fitted by a function of $\exp{(-t/\tau)^{0.3}}$ in comparison with the simple exponential decay for the whisker, as highlighted in Fig. 3 \cite{comment}. These facts suggest that the type of phase transition for ring-shaped CDWs is essentially different from that for whisker. In the slow relaxation expressed by a power or stretched exponential function, the enhancement of phase fluctuations originating from the ring topology must play an important role.

\begin{figure}[t]
\begin{center}
\includegraphics[width=85mm]{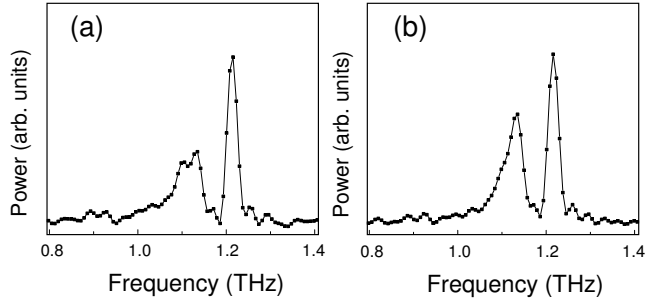}
\caption{Fourier transform of the oscillation signal obtained by subtracting the exponential contributions from the data: (a) spectrum for the whisker at 28 K and (b) spectrum for the ring at 27 K. The peak at lower frequency is identified as that of an amplitude mode (AM).}
\end{center}
\end{figure}

Nevertheless, it is also possible to explain the different transient dynamics in terms of the difference in the crystal inhomogeneities between the samples. Since the phase fluctuation is sensitive to impurities, dislocations, and other disorders, we cannot completely exclude the contribution of the crystal inhomogeneities to the present results. As a simple check, we evaluated a coherent oscillation of CDWs induced by an instantaneous optical excitation. Since the dephasing time of collective motion is determined by the CDW coherent length, we can qualitatively compare the crystal inhomogeneities.

Figures 4(a) and (b) show the fast Fourier transform spectra of the ring and the whisker, respectively, at almost the same temperature. In each spectrum, two distinct peaks are clearly visible. The lower mode at around 1.1 THz shows a softening behavior with increasing temperature and completely disappears above {\it T$_{c2}$}, indicating collective excitation of the CDW (amplitude mode: AM) associated with the incommensurate phase of {\it T$_{c2}$}, while the higher mode is identified as a coherent phonon because of its less pronounced temperature dependence. Note that the AM mode in NbSe$_3$ has not been observed experimentally so far. However, its frequency and temperature dependence are qualitatively similar to the AM oscillation observed in several other CDW compounds \cite{Demsar, AM}. The linewidths of AM obtained from the Lorentzian fits to the spectra are $\sim$ 75 GHz for the whisker and $\sim$ 48 GHz for the ring, which are comparable to each other. We thus believe that the contribution of the crystal inhomogeneities to the phase fluctuation is nearly the same. 

In summary, we have investigated the ultrafast SP dynamics in ring and whisker NbSe$_3$ crystals, from the viewpoint of the crystal topology. In the phase transition around {\it T$_{c2}$}, the temperature dependence of the SP recombination time observed in the ring is quantitatively and qualitatively different from that in the whisker: quenching of the divergence of $\tau_s$ and discrepancy from BCS-like temperature dependence. These results can be explained in terms of the enhanced phase fluctuation in the closed-loop topology.

Authours sincerely acknowledge Dr. T. Nogawa and Dr. K. Nemoto for their fruitful discussions. This work is supported by a Grant-in-Aid for the 21st Century COE program "Topological Science and Technology".



\begin{references}
\bibitem{Gruner} G. Gr\"{u}ner , Rev. Mod. Phys. \textbf{60}, 1129 (1988);
{\it Density Waves in Solids} (Addison-Wesley , Reading, MA, 1994).
\bibitem{size} E. Slot, M. A. Holst, H. S. J. van der Zant, and S. V. Zaitsev-Zotov, Phys. Rev. Lett. \textbf{93}, 176602 (2004); J. McCarten, D. A. DiCarlo, M. P. Maher, T. L. Adelman, and R. E. Thorne, Phys. Rev. B \textbf{46}, 4456 (1992); J. C. Gill, Synthetic Met. \textbf{43}, 3917 (1991)
%
\bibitem{Tanda} S. Tanda, T. Tsuneta, Y. Okajima, K. Inagaki, K. Yamaya, N. Hatakenaka, Nature \textbf{417}, 397 (2002).
\bibitem{Sweetland}E. Sweetland, C. Y. Tsai, B. A. Wintner, J. D. Brock and R. E. Thorne, Phys. Rev. Lett. \textbf{65}, 3165 (1990). 
\bibitem{Tsuneta} T. Tsuneta, and S. Tanda, J. Crystal Growth \textbf{264}, 223 (2004); T. Tsuneta, S. Tanda, Y. Okajima, K. Inagaki, K. Yamaya, Physica B \textbf{329}, 1544 (2003).
%
\bibitem{Demsar} J. Demsar, K. Biljakovic, and D. Mihailovic, Phys. Rev. Lett. \textbf{83}, 800 (1999).
\bibitem{AM} J. Demsar, L. Forro, H. Berger, and D. Mihailovic, Phys. Rev. \textbf{B 66}, 041101 (2002); Y. Toda, K. Tateishi, and S. Tanda, Phys. Rev. \textbf{B 70}, 033106 (2004); Y. Ren, G. L\"{u}pke, and Z. Xu, Appl. Phys. Lett. \textbf{84}, 2169 (2004).
\bibitem{Kabanov1} V.V. Kabanov, J. Demsar, B. Podobnik, and D. Mihailovic, Phys. Rev. B 59, 1497 (1999). 
\bibitem{Kabanov2} V.V. Kabanov, J. Demsar, and D. Mihailovic, Phys. Rev. B 61, 1477 (2000).
%
\bibitem{STM} A. Fournel, J.P. Sorbier, M. Konczykowski, and P. Monceau, Phys. Rev. Lett. \textbf{57}, 2199 (1986); Z. Dai, C.G. Slough, and R.V. Coleman, Phys. Rev. Lett. \textbf{66}, 1318 (1991); A.A. Sinchenko and P. Monceau, Phys. Rev. B \textbf{67}, 125117 (2003).
\bibitem{ARPES} J. Sch\"{a}fer, M. Sing, R. Claessen, Eli Rotenberg, X.J. Zhou, R.E. Thorne, and S.D. Kevan, Phys. Rev. Lett. \textbf{91}, 066401 (2003).
\bibitem{Nogawa} T. Nogawa, and K. Nemoto, unpublished.
\bibitem{comment} This suggests another possibility that $\tau_s$ for the ring is underestimated by fitting the normal exponential function and thus shows no singular behavior at {\it T$_{c2}$}.
\end{references}
\end{document}